\documentclass[conference]{IEEEtran}

\usepackage{hyperref}
\hypersetup{
     colorlinks = true,
     citecolor = green,
     }

\ifCLASSINFOpdf
  \usepackage[pdftex]{graphicx}
\else
\fi

\usepackage{soul}
\usepackage{amsmath}
\usepackage{listings}
\usepackage{xcolor} 

\usepackage{listings}
\usepackage{xcolor}

\definecolor{cs-blue}{rgb}{0.035, 0.114, 0.235}
\definecolor{turq}{rgb}{0.365, 0.835, 0.851}

\usepackage{tcolorbox}

\tcbset{
  takeawaybox/.style={
    colback=turq!7!white, 
    colframe=turq, 
    coltitle=black, 
    fonttitle=\bfseries\small, 
    title=Key Takeaways: 
  }
}

\usepackage{subcaption}

\usepackage{framed}
\usepackage{balance}
\balance

\begin{document}
%
\title{Ghost Echoes Revealed: Benchmarking Maintainability Metrics and Machine Learning Predictions Against Human Assessments}


%
\author{\IEEEauthorblockN{Markus Borg\IEEEauthorrefmark{1}, Marwa Ezzouhri\IEEEauthorrefmark{2}, Adam Tornhill\IEEEauthorrefmark{3}}
\IEEEauthorblockA{\IEEEauthorrefmark{1}CodeScene, Malmö, Sweden and Lund University, Lund, Sweden, markus.borg@codescene.com}
\IEEEauthorblockA{\IEEEauthorrefmark{2}University of Clermont Auvergne, Clermont-Ferrand, France, marwa.ezzouhri@etu.uca.fr}
\IEEEauthorblockA{\IEEEauthorrefmark{3}CodeScene, Malmö, Sweden, adam.tornhill@codescene.com}
}


\maketitle

\begin{abstract}
As generative AI is expected to increase global code volumes, the importance of maintainability from a human perspective will become even greater. Various methods have been developed to identify the most important maintainability issues, including aggregated metrics and advanced Machine Learning (ML) models. This study benchmarks several maintainability prediction approaches, including State-of-the-Art (SotA) ML, SonarQube's Maintainability Rating, CodeScene's Code Health, and Microsoft's Maintainability Index. Our results indicate that CodeScene matches the accuracy of SotA ML and outperforms the average human expert. Importantly, unlike SotA ML, CodeScene also provides end users with actionable code smell details to remedy identified issues. Finally, caution is advised with SonarQube due to its tendency to generate many false positives. Unfortunately, our findings call into question the validity of previous studies that solely relied on SonarQube output for establishing ground truth labels. To improve reliability in future maintainability and technical debt studies, we recommend employing more accurate metrics. Moreover, reevaluating previous findings with Code Health would mitigate this revealed validity threat.
\end{abstract}


%

\section{Introduction} \label{sec:intro}
Maintainability remains an elusive goal for many software organizations. Despite numerous studies highlighting the costs of poor code quality and \textit{Technical Debt} (TD), organizations routinely prioritize new feature development over refactoring~\cite{reboucas_de_almeida_aligning_2018,ampatzoglou_financial_2015,tornhill_code_2022}. To tackle maintainability challenges at scale, reliable metrics are needed to quantify potential issues. We believe the challenges will be exacerbated with the advent of widespread AI-assisted coding practices, i.e., human developers will have an even harder time staying on top of codebases in the future. 
Reliable metrics that can quantify potential issues are essential. Such metrics not only help in identifying and prioritizing files that need improvement based on development activity~\cite{tornhill_prioritize_2018}, but also facilitate the setting of explicit targets for enhancements~\cite{borg_quality_2024}.

Numerous metrics have been proposed to evaluate various aspects of source code quality. A systematic review by Nu\~{n}ez-Varela \textit{et al.} identified almost 300 metrics in papers published between 2010 and 2015. The authors show that decades-old metrics such as \textit{Lines of Code} (LoC), \textit{McCabe's Cyclomatic Complexity} (CC), and \textit{Halstead Volume} (HV) remain popular in research papers. Beyond these, various object-oriented metrics dominated the primary studies included from the early 2010s, including the established Chidamber and Kemerer set~\cite{chidamber_metrics_1994}. Unfortunately, research suggests that the correlation between maintainability and individual metrics such as LoC, CC, and HV is weak at best~\cite{peitek_program_2021}.

To remedy the limitations of individual metrics, several approaches to combine them have been proposed. To support this line of research, i.e., maintainability prediction of source code files, Schnappinger \textit{et al.} (2020) presented a manually annotated \textit{Maintainability Dataset} (MainData) containing 519 files, including 304 open-source files ~\cite{schnappinger_defining_2020}. Using this dataset, Schnappinger \textit{et al.} later trained accurate \textit{Machine Learning} (ML) models that largely match the human labels~\cite{schnappinger_human-level_2021}. Two years later, Bertrand \textit{et al.} trained ML models that outperformed these results~\cite{bertrand_replication_2023} on MainData's open-source part -- an ensemble model using \textit{Adapted Boosting} (AdaBoost) now constitutes the \textit{State-of-the-Art} (SotA) for binary maintainability prediction on MainData (see  Section~\ref{sec:datacoll} for details). However, it remains unclear to what extent models trained on such a small set of manually annotated files generalize to other projects.

Instead of training ML models, individual code metrics can be aggregated using algorithmic models into single-number maintainability scores. As no training is needed, algorithmic models can be applied out-of-the-box to any software project implemented in a supported programming language. Oman and Hagemeister proposed the Maintainability Index in 1992, which combines LoC, CC, HV, and comment density into a single number~\cite{oman_metrics_1992}. Several contemporary tools implement various maintainability scores. Microsoft implements a version of the \textit{Maintainability Index} (MS-MI) scaled within a range of 1-100 in their products. SonarQube translates rule violations into a Maintainability Rating between A--E. Additionally, CodeScene provides the Code Health metric, aggregating a set of code smells into a score between 1 and 10. But how do these algorithmic models perform relative to the SotA ML model? 

In this study, we conduct benchmarking~\cite{hasselbring_benchmarking_2021} to compare three algorithmic maintainability scores used in industry tools to the SotA ML model. We collectively refer to the set of competing alternatives as prediction approaches. First, we reproduce Bertrand \textit{et al.}'s results for binary maintainability classification on MainData. Subsequently, we compare the results against the out-of-the-box maintainability verdicts provided by SonarQube, MS, and CodeScene. Moreover, we compare the results against a naïve rule-based LoC baseline. For this part of the study, we rely on the standard evaluation metrics: accuracy, precision, recall, and F1-score.

In line with previous work, the evaluation metrics in the first part assess how well the prediction approaches identify files that belong to the majority of maintainable files -- we refer to this as \textit{Use Case} 1 (UC1) as ``Maintainability Prediction.'' However, we argue that it is more important for tools to correctly identify the minority of unmaintainable files. This viewpoint is better aligned with quality assurance tools in general, such as code smell detectors and vulnerability scanners. Thus, we also provide evaluation metrics corresponding to this opposite view -- we refer to this as UC2 ''Liability Prediction'' -- and focus on the F0.5-score, which penalizes false positives.

Second, we perform a comprehensive analysis of the predictive power of the industrial tools' underlying maintainability metrics. We do this by considering the tools as binary classifiers and varying their internal thresholds used to determine whether a file is maintainable or not. We present the results as \textit{Receiver Operating Characteristic} (ROC) curves and calculate the \textit{Areas Under the Curves} (AUC) to draw conclusions.  Furthermore, we again compare the maintainability metrics against SotA ML and the LoC baseline. The use of AUC is particularly appropriate here, as it provides a robust single-value summary of the prediction approaches' discriminative abilities.

Our research is driven by two primary research questions:
\begin{itemize}
\item[RQ1] How do contemporary industrial tools' maintainability assessments compare to SotA ML in matching human experts' judgments?
\item[RQ2] What is the comparative predictive power of the tools' underlying maintainability metrics relative to SotA ML?
\end{itemize}

Our findings show that CodeScene matches SotA ML and outperforms the average human expert in terms of F1-score. Meanwhile, MS-MI demonstrates moderate accuracy, while SonarQube generates many false positives -- readers should beware of its ghost echoes. Furthermore, Bertrand~\textit{et al.}~\cite{bertrand_replication_2023} remains SotA on MainData, slightly surpassing the predictive power of Code Health as shown by the corresponding AUC scores. On the other hand, maintainability issues identified using metrics based on code smells, as provided by CodeScene and SonarQube, include actionable remediation steps. We provide a complete replication package on Zenodo~\cite{borg_ghost_2024}.

The rest of the paper is organized as follows. Section~\ref{sec:bg} introduces how the different maintainability scores evaluated in this study are calculated. The section also describes how MainData was created in previous work. Section~\ref{sec:rw} positions our study in light of related work. In Section~\ref{sec:method}, we explain the research method, and the results follow in Section~\ref{sec:res}. We discuss the major threats to validity and implications for research and practice in Sections~\ref{sec:threats} and~\ref{sec:impl}, respectively. Finally, we conclude the paper in Section~\ref{sec:conc}.

\section{Background} \label{sec:bg}
Numerous papers have put forward metrics to evaluate source code quality, resulting in an extensive array of options in this field~\cite{nunez-varela_source_2017}. In this section, we present how three contemporary industrial solutions address maintainability aspects: the MS-MI, SonarQube, and CodeScene. Finally, we briefly describe the credible creation of MainData.

\subsection{Microsoft's Maintainability Index}
Oman and Hagemeister proposed a seminal maintainability index in 1992~\cite{oman_metrics_1992}, combining LoC, CC, HV, and the percentage of code comments. Large projects from Hewlett-Packard were used to calibrate constants for a formula that combined the four code features. In the 80s and 90s, this approach to developing algorithmic models -- such as the COCOMO suite~\cite{boehm_cost_1995} -- was popular in software engineering research.

In 2011, MS integrated a simplified version of this index into Visual Studio, excluding the density of code comments. In this process, they also revised the formula to bound the result within the range 0--100 as follows:

\begin{align}
\text{MS-MI} &= \max\left(0, \right. \nonumber \\
&\quad \left. \left(171 - 5.2 \times \ln(HV) - 0.23 \times CC - \right. \right. \nonumber \\
&\quad \left. \left. 16.2 \times \ln(LoC)\right) \times \frac{100}{171}\right)
\end{align}

MS claims to have set conservative thresholds to avoid false positives~\cite{microsoft_code_nodate}. Files are categorized into three intervals based on their scores: Red (0--9), Yellow (10-19), and Green (20-100).

\subsection{SonarQube, SQALE, and Maintainability Rating}
SonarQube is the most commonly used advanced static code analysis tool in industry~\cite{lenarduzzi_technical_2021}. Developers can run SonarQube as a cloud service or the tool can be installed locally for execution on-premise. SonarQube's quality model considers three dimensions: reliability, maintainability, and security. For each dimension, the tool has defined rules that trigger SonarQube issues if violated. Moreover, each rule is connected to one of five severity levels: BLOCKER, CRITICAL, MAJOR, MINOR, and INFO. Details about the SonarQube rules are available in the documentation~\cite{sonarqube_sonarqube_nodate}.

SonarQube uses the SQALE method to convert rule violations to estimates of TD remediation costs~\cite{letouzey_sqale_2012}. The SQALE method defines a \textit{quality model} that breaks down non-functional code quality requirements into specific code rules. Subsequently, an \textit{analysis model} provides a conversion from rule violations to the cost of actions needed to correct them. The costs are expressed in development time. The total \textit{TD remediation Time} (TD Time) for a file is the sum of all estimated development times associated with the independent rule violations. Finally, the SQALE method calculates the ratio between TD Time and the estimated total development time of the file --- the latter estimated as 0.06 days per LoC. For each file, this ratio is calculated, i.e., TD Ratio.

While SonarQube uses the SQALE method, there are some variation points in the implementation compared to the original 2012 paper by Letouzey~\cite{letouzey_sqale_2012}. First, SonarQube has evolved its own set of maintainability rules. Second, SonarQube's has evolved thresholds for its Maintainability Rating (A--E) as follows:

\begin{itemize}
    \item[A:] TD Ratio $\leq 5\%$
    \item[B:] 5\% $<$ TD Ratio $\leq$ 10\%
    \item[C:] 10\% $<$ TD Ratio $\leq$ 20\%
    \item[D:] 20\% $<$ TD Ratio $\leq$ 50\%
    \item[E:] TD Ratio $>$ 50\%
\end{itemize}

\subsection{CodeScene and Code Health}
CodeScene is a software engineering intelligence platform that combines code quality analysis with people and organizational perspectives. By connecting code with the humans developing and maintaining it, the solution can more accurately prioritize where refactoring efforts would provide a positive return on investment. Moreover, CodeScene can estimate if the main bottlenecks for the software development organization rather reside in poor knowledge distribution or inadequate team structures.  

CodeScene measures code quality using its Code Health metric~\cite{codescene_code_nodate}. Code Health focuses on how cognitively difficult it is for human developers to comprehend what the code is doing. The metric aligns with the mindset that the best strategy for gauging code quality is to aggregate a set of specific complexity attributes~\cite{fenton_software_1994}. CodeScene parses source code to identify the presence of established code smells, e.g., God Class, God Methods, and Duplicated Code~\cite{lacerda_code_2020}. CodeScene currently supports 31 languages. Most of the code smells it identifies are language agnostic, differing only in their threshold settings across languages.

For Java, which is the focus of this paper, CodeScene detects 25 code smells. The presence of code smells is combined into a numeric value between 1 and 10 -- the lower end represents extremely poor maintainability, and the upper end indicates top-notch code that is easy to maintain. CodeScene categorizes files into one of three sub-intervals: healthy (9 or higher), warning (between 4 and 9), and alert (lower than 4).

\subsection{The Maintainability Dataset} \label{sec:maindata}
In 2020, Schnappinger \textit{et al.} made a significant contribution to maintainability research as they published MainData~\cite{schnappinger_defining_2020}. Relying on a sophisticated manual annotation process, they created the most reliable ground truth maintainability labels available to date. Seventy professional developers contributed about 2,000 manual maintainability assessments of nine Java projects. In this study, we use the labels on 304 files from MainData's five open-source projects:

\begin{itemize}
    \item ArgoUML: UML modeling tool.
    \item Art of Illusion: 3D modeling and rendering studio.
    \item Diary Management: multi-user calendar tool.
    \item JUnit 4: unit testing framework.
    \item JSweet: Java to JavaScript/TypeScript transpiler.
\end{itemize}

Each file has been evaluated by at least three human experts, who assigned a label based on a four-point ordinal scale. This scale measures readability, understandability, complexity, modularization, and overall maintainability. In line with prior studies, we use the majority opinion on the overall label as the ground truth. For binary prediction, files that receive the higher two values on this scale are considered maintainable. MainData is imbalanced with 78\% maintainable files.

Schnappinger \textit{et al.}'s work on creating a ground truth for maintainability judgments reveals substantial variability among the annotations by human experts. For 17\% of the source code files, the experts disagreed substantially. To provide a baseline for future research, the researchers established the ``average human expert'' baseline by comparing individual ratings with a consolidated consensus for each source code file. This baseline, used for comparisons in Table~\ref{tab:res}, reflects the average agreement level between human experts and a collective consensus. With an F1-score of 0.88 for maintainability prediction, it is evident that the task is subjective and non-trivial.

\section{Related Work} \label{sec:rw}
This section introduces related work on 1) training ML models for maintainability prediction and 2) empirical evaluations of maintainability metrics.

\subsection{Supervised learning for maintainability prediction}
Many researchers have proposed training ML models for predicting how maintainable a file is. Restricted to object-oriented systems, Alsolai and Roper identified 56 primary studies in a systematic literature review~\cite{alsolai_systematic_2020}. They report that training ensemble models trained on low-level product (code) metrics have obtained the best results. Furthermore, they highlight a growing demand for publicly available datasets. Schnappinger \textit{et al.} responded to this call and created MainData as described in Section~\ref{sec:maindata}.

Schnappinger \textit{et al.} later trained maintainability prediction models using the manually annotated data~\cite{schnappinger_human-level_2021}. The models were trained on a set of features extracted from a variety of static code analysis tools, i.e., ConQAT, Designite, SD Metrics, SonarQube, Sourcemeter, and Teamscale. Out of the 132 potentially useful features, the best-performing models (F1=0.91 and AUC=0.82) were trained using only five, indicating that many features either lacked predictive power or were highly correlated. For binary maintainability classification, the authors' ML models slightly outperformed the average human expert (F1=0.88 and AUC=0.83).

Bertrand \textit{et al.} replicated Schnappinger \textit{et al.}'s study and obtained more accurate maintainability predictions~\cite{bertrand_replication_2023}. Moreover, they provide a substantially more transparent replication package with a mature pipeline for model training and validation. In contrast to the original study, the replication publishes 34 raw features for all files, extracted using their own open-source tool Javanalyzer~\cite{bertrand_building_2022}. For binary maintainability classification, Bertrand \textit{et al.}'s best-performing ensemble model now constitutes SotA on MainData (F1=0.95 and AUC=0.97). In this study, we refer to this as SotA ML and use it for comparisons. The details of the SotA ML configuration are further described in Section~\ref{sec:datacoll}.

Both Schnappinger \textit{et al.} and Bertrand \textit{et al.} extend their analysis beyond binary prediction, focusing primarily on ordinal prediction. They argue that binary prediction oversimplifies the complexity of the maintainability quality, and instead focus on predicting MainData's labels on the four-point ordinal scale. In this paper, we limit our evaluation to binary prediction as it simplifies the comparison and facilitates clearer communication of the results to practitioners. However, maintainability metrics are inherently finer-grained than ordinal predictors as they provide scores in the range from abysmal to excellent levels. Still, binary prediction using industrial tools' default thresholds for the corresponding metrics is critical to study. Surpassing these thresholds serves as a call to action for developers.

\subsection{Empirical evaluations of maintainability metrics}
Several researchers have conducted empirical evaluations of SonarQube metrics, often in the context of TD. Lenarduzzi and colleagues have been particularly active in scrutinizing the accuracy of SonarQube. They have found that the correlation between the number of SonarQube rule violations in a file is not correlated with the time taken to fix issues~\cite{lenarduzzi_technical_2021}. In the same vein, they have found that SonarQube's TD Time is inaccurate and generally overestimated~\cite{saarimaki_accuracy_2019,baldassarre_diffuseness_2020}. In a comparison of six static code analysis tools, they found that SonarQube had the lowest precision (0.18) on a manually-defined ground truth~\cite{lenarduzzi_critical_2023}. 

Lenarduzzi \textit{et al.} have also made an important contribution toward maintainability benchmarking by publishing the TD dataset~\cite{lenarduzzi_technical_2019}. It contains fine-granular commit data from 33 Java projects from the Apache Software Foundation. However, the massive dataset does not provide any human labels. Instead, despite its questionable accuracy, SonarQube output is used as a maintainability proxy. Another example of a recent TD study relying on SonarQube output is Paudel \textit{et al.}~\cite{paudel_towards_2024}. Despite all critical studies, SonarQube might still be a useful tool for binary maintainability prediction -- which we investigate in this study.

We have previously evaluated Code Health in three studies based on large-scale proprietary projects. First, we demonstrated that files within the alert interval contain 15 times more defects compared to healthy ones~\cite{tornhill_code_2022}. The study also found a velocity impact as developers need 124\% more time when making changes to such files. Further, unhealthy files were found to exhibit considerably greater variability in task completion times. Second, studying a bigger dataset, we showed a practically significant file-level association between Code Health, defect counts, and issue resolution times~\cite{borg_increasing_2024}. Third, we found that developers with low file ownership require more time to resolve issues
in alert-level code~\cite{borg_u_2023}. Based on our previous results, we believe that Code Health can successfully be used for maintainability prediction.

Thanks to MainData's reliable ground truth labels, we proceed by comparing the accuracy of the industrial tools SonarQube and CodeScene. To provide a richer picture, we compare the maintainability predictions with SotA ML, MS-MI, and a simple LoC baseline.

\section{Research Method} \label{sec:method}
This section describes the details of the empirical study.

\subsection{Data collection and preparation} \label{sec:datacoll}
We downloaded MainData from figshare~\cite{schnappinger_defining_2020} and immediately noticed that class files for the included project ArgoUML were missing. We reached out to the creators of the dataset who confirmed the absence. Unfortunately, the original authors no longer had access to the missing files. Since SonarQube requires class files to analyze Java projects, we had to resolve this issue.

ArgoUML is an old project that started already in 1998. Schnappinger \textit{et al.} claimed to have analyzed a 2010 build of ArgoUML~\cite{schnappinger_defining_2020}, but it is unclear exactly which version it was. Since then, ArgoUML has migrated from SourceForge to GitHub and from Maven builds to GitHub Actions. We could not build any 2010 versions since dependencies were missing and no longer hosted online. Instead, we built a more recent version, commit 6969f51 from 2020, and verified that the 74 relevant ArgoUML source code files are identical to the versions in MainData. Note that ArgoUML has not been actively developed since it migrated to GitHub. We also contacted the lead maintainer of ArgoUML, who confirmed that no changes to the product code had been made between 2010 and 2020.

We downloaded Bertrand \textit{et al.}'s replication package~\cite{bertrand_replication_2023}, containing the code used to train their classifiers and the matching training data. To ensure we used the preprocessing choices and hyperparameters corresponding to the best results presented in the paper (SotA ML), we confirmed our configuration\footnote{The source code is available in our replication package~\cite{borg_ghost_2024}.} with the original authors. The authors obtained the best results with an AdaBoost classifier~\cite{freund_experiments_1996}, an ensemble approach that combines several weak models to create an accurate model. We successfully reproduced the best results for UC1 using an AdaBoost classifier implemented in scikit-learn, configured with 150 estimators and a learning rate of 0.5, trained on 33 low-level features.

\subsection{Data analysis} \label{sec:dataanal}
We analyzed the five MainData projects with the default settings of CodeScene Community Edition 6.4.13 and SonarQube Community Edition Version 10.0 (build 68432). The MS-MI was calculated using the IntelliJ IDEA plugin MetricsTree v.2024.1.0\footnote{\url{https://plugins.jetbrains.com/plugin/13959-metricstree}}. These three tools provide deterministic output, i.e., the values for Code Health, TD Ratio, TD Time, and MS-MI do not change across runs. On the other hand, Bertrand \textit{et al.}'s implementation of SotA ML uses shuffled stratified 5-fold cross-validation~\cite{bertrand_replication_2023} to allow robust comparisons. We reuse this setup and the reported results for SotA ML correspond to an average of the corresponding runs. 

For UC1, we use the same binary classification setup as in previous work on MainData~\cite{schnappinger_human-level_2021,bertrand_replication_2023}. The ground truth for each file is that it is considered \textit{maintainable} if the human annotators' majority voting resulted in an overall score of 0 or 1 on the four-point ordinal scale. Otherwise, the file is considered \textit{unmaintainable}. For UC2, as presented in Section~\ref{sec:intro}, we invert the ground truth. Consequently, this leads to a more common discussion of the four cells in the confusion matrix suitable for a detection system. Thus, for UC2, a true positive is a file that is correctly predicted as unmaintainable, whereas a false positive is a file that is incorrectly highlighted as unmaintainable. In this realistic context, we believe that false positives are a bigger problem than false negatives.

The seven binary prediction approaches under study for RQ1 are the following:

\begin{itemize}
    \item[(A)] SotA ML uses supervised ensemble learning, AdaBoost, to train a classifier on MainData. We reproduce previous results~\cite{bertrand_replication_2023} and report mean results from cross-validation.
    \item[(B)] CodeScene's Code Health is an algorithmic approach that relies on penalizing files by aggregating high-level code smells. A file is considered maintainable if its Code~Health is 9.0 or higher.
    \item[(C)] Baseline LoC is a simple but accurate baseline from previous work~\cite{bertrand_replication_2023}. If a file contains no more than 275~LoC, it is considered maintainable.
    \item[(D)] The MS-MI is an algorithmic approach that combines three low-level code metrics into a score between 1 and 100. A file is considered maintainable if the index is 20 or higher.
    \item[(E)] SonarQube implements an algorithmic approach that aggregates hundreds of rule violations and low-level code smells. TD Time is not used for maintainability prediction in SonarQube, but since it outperforms TD Ratio we choose to report it. For MainData, the best results are obtained when a file is considered maintainable if its TD Time is less than 189 minutes.
    \item[(F)] The performance of the average human expert for UC1 on MainData as reported by Schnappinger \textit{et al.}~\cite{schnappinger_human-level_2021}. The authors reported that the average human agreed with the consensus in 70\% of the cases, which highlights that maintainability is a subjective quality.
    \item[(G)] TD Ratio is the out-of-the-box maintainability prediction approach provided by SonarQube, resulting in a Maintainability Rating A--E. A file is considered maintainable if its TD Ratio is no more than 0.05.
\end{itemize}

To provide a robust comparison of the maintainability prediction approaches for RQ2, we investigate ROC curves. For Code Health, we plot what a rule-based binary classifier would yield for different threshold values. The default value is 9.0, but we illustrate the entire interval 1-10 with a step size of 0.1. We follow the same procedure for SonarQube, i.e., we plot the results of rule-based binary classifiers using threshold values for TD Ratio and TD Time. The stepsizes are 0.005 and 5 min, respectively. Analogously, we plot a ROC curve for MS-MI representing different thresholds from 0 to 100 with a stepsize of 1. Finally, we plot a curve also for the LoC baseline with a stepsize of 10 LoC. 

We report a set of standard performance metrics. \textit{Accuracy} (Acc) is the fraction of correctly predicted files. AUC, a robust single-value metric for binary classifiers, is calculated using NumPy's implementation of the composite trapezoidal rule to estimate the area under the curve. Both Acc and AUC remain the same for both UC1 and UC2, as the discriminative powers of the prediction approaches do not change with the inverted ground truth. The following metrics, however, will be reported per UC. \textit{Precision} (Pr) is the proportion of highlighted files that are correctly predicted. \textit{Recall} (Rc) is the proportion of files that were correctly highlighted. The \textit{F1-score} (F1) is the harmonic mean of Pr and Rc. Since false positives should be avoided for UC2, we focus on the F0.5-score (F0.5) that weighs Pr twice as heavily as Rc. 

\section{Results and Discussion} \label{sec:res}
Table~\ref{tab:res} presents the results of the different prediction approaches ordered by AUC scores in the rightmost column. Column three shows which features are used by the approach. Column four lists how the features are aggregated into a prediction or score. Column five shows the thresholds used for binary classification, using tool defaults when possible. The performance metrics follow in the remaining columns, with results separated for UC1 (Maintainability Prediction) and UC2 (Liability Prediction).

\begin{table*}[h]
\caption{Performance of the prediction approaches ordered by AUC scores. Rows in italic font highlight the performance of three contemporary industrial tools.}
\label{tab:res}
\begin{tabular}{clllcc|ccc|cccc|c}
\cline{7-13}
                                  &                                                        &                                            &                                           & \multicolumn{1}{l}{}                    &               & \multicolumn{3}{c|}{\textbf{UC1: Main. Pred.}}                                          & \multicolumn{4}{c|}{\textbf{UC2: Liab. Pred.}}                                                                               &                                    \\ \hline
\multicolumn{1}{|c|}{\textbf{ID}} & \multicolumn{1}{l|}{\textbf{Prediction Approach}}      & \multicolumn{1}{l|}{\textbf{Features}}     & \multicolumn{1}{l|}{\textbf{Aggregation}} & \multicolumn{1}{c|}{\textbf{Threshold}} & \textbf{Acc}  & \multicolumn{1}{c|}{\textbf{Pr}}   & \multicolumn{1}{c|}{\textbf{Rc}}   & \textbf{F1}   & \multicolumn{1}{c|}{\textbf{Pr}}   & \multicolumn{1}{c|}{\textbf{Rc}}   & \multicolumn{1}{c|}{\textbf{F1}}   & \textbf{F0.5} & \multicolumn{1}{c|}{\textbf{AUC}}  \\ \hline
\multicolumn{1}{|c|}{(A)}           & \multicolumn{1}{l|}{SotA ML}                           & \multicolumn{1}{l|}{Code metrics}          & \multicolumn{1}{l|}{AdaBoost}             & \multicolumn{1}{c|}{0.5}                & 0.92          & \multicolumn{1}{c|}{0.95}          & \multicolumn{1}{c|}{0.95}          & 0.95          & \multicolumn{1}{c|}{0.83}          & \multicolumn{1}{c|}{0.81}          & \multicolumn{1}{c|}{0.82}          & 0.82          & \multicolumn{1}{c|}{0.97}          \\ \hline
\multicolumn{1}{|c|}{\textit{(B)}}  & \multicolumn{1}{l|}{\textit{CodeScene Code Health}}    & \multicolumn{1}{l|}{\textit{Code smells}}  & \multicolumn{1}{l|}{\textit{Algorithmic}} & \multicolumn{1}{c|}{\textit{9.0}}       & \textit{0.93} & \multicolumn{1}{c|}{\textit{0.94}} & \multicolumn{1}{c|}{\textit{0.97}} & \textit{0.96} & \multicolumn{1}{c|}{\textit{0.89}} & \multicolumn{1}{c|}{\textit{0.77}} & \multicolumn{1}{c|}{\textit{0.83}} & \textit{0.87} & \multicolumn{1}{c|}{\textit{0.95}} \\ \hline
\multicolumn{1}{|c|}{(C)}           & \multicolumn{1}{l|}{LoC}                               & \multicolumn{1}{l|}{LoC only}              & \multicolumn{1}{l|}{None}                 & \multicolumn{1}{c|}{275}                & 0.91          & \multicolumn{1}{c|}{0.92}          & \multicolumn{1}{c|}{0.97}          & 0.95          & \multicolumn{1}{c|}{0.87}          & \multicolumn{1}{c|}{0.71}          & \multicolumn{1}{c|}{0.78}          & 0.83          & \multicolumn{1}{c|}{0.95}          \\ \hline
\multicolumn{1}{|c|}{\textit{(D)}}  & \multicolumn{1}{l|}{\textit{MS Maintainability Index}} & \multicolumn{1}{l|}{\textit{Code metrics}} & \multicolumn{1}{l|}{\textit{Algorithmic}} & \multicolumn{1}{c|}{\textit{20}}        & \textit{0.84} & \multicolumn{1}{c|}{\textit{0.88}} & \multicolumn{1}{c|}{\textit{0.92}} & \textit{0.90} & \multicolumn{1}{c|}{\textit{0.64}} & \multicolumn{1}{c|}{\textit{0.55}} & \multicolumn{1}{c|}{\textit{0.59}} & \textit{0.62} & \multicolumn{1}{c|}{\textit{0.89}} \\ \hline
\multicolumn{1}{|c|}{(E)}           & \multicolumn{1}{l|}{SonarQube TD Time}                 & \multicolumn{1}{l|}{Code smells}           & \multicolumn{1}{l|}{Algorithmic}          & \multicolumn{1}{c|}{189}                & 0.86          & \multicolumn{1}{c|}{0.87}          & \multicolumn{1}{c|}{0.95}          & 0.91          & \multicolumn{1}{c|}{0.75}          & \multicolumn{1}{c|}{0.50}          & \multicolumn{1}{c|}{0.60}          & 0.68          & \multicolumn{1}{c|}{0.86}          \\ \hline
\multicolumn{1}{|c|}{(F)}           & \multicolumn{1}{l|}{Average human expert}              & \multicolumn{3}{c|}{Expert opinion}                                                                                              & 0.70          & \multicolumn{1}{c|}{0.88}          & \multicolumn{1}{c|}{0.88}          & 0.88          & \multicolumn{1}{c|}{--}            & \multicolumn{1}{c|}{--}            & \multicolumn{1}{c|}{--}            & --            & \multicolumn{1}{c|}{0.83}          \\ \hline
\multicolumn{1}{|c|}{\textit{(G)}}  & \multicolumn{1}{l|}{\textit{SonarQube TD Ratio}}       & \multicolumn{1}{l|}{\textit{Code smells}}  & \multicolumn{1}{l|}{\textit{Algorithmic}} & \multicolumn{1}{c|}{\textit{0.05}}      & \textit{0.61} & \multicolumn{1}{c|}{\textit{0.75}} & \multicolumn{1}{c|}{\textit{0.74}} & \textit{0.75} & \multicolumn{1}{c|}{\textit{0.11}} & \multicolumn{1}{c|}{\textit{0.12}} & \multicolumn{1}{c|}{\textit{0.12}} & \textit{0.12} & \multicolumn{1}{c|}{\textit{0.60}} \\ \hline
\end{tabular}
\end{table*}

Figures~\ref{fig:roc_maint} and~\ref{fig:roc_liab} present ROC curves for six prediction approaches corresponding to UC1 and UC2, respectively. Note that the approaches' AUC scores remain the same for UC1 and UC2 despite different ROC curves. This reflects that the discriminative powers of the binary classifiers are the same, but the meanings of the True Positive Rates (TPR) and False Positive Rates (FPR) are different. Finally, the stars on the curves for (B), (D), and (G) indicate the prediction performance corresponding to the thresholds implemented in the respective tools, i.e., Code Health 9.0, MS-MI 20, and TD Ratio 0.05. Next, we answer the two research questions.

\subsection{RQ1: Industrial Tools, SotA ML, and Human Experts} \label{sec:res_rq1}
For UC1, we find that all prediction approaches except SonarQube TD Ratio outperform an average human expert (F) on MainData. The task of identifying maintainable files, i.e., the 78\% subset, leads to high F1 scores. SotA ML (A), Code Health (B), and the LoC baseline (C) reach about 0.95. TD Time (E) and MS-MI (D) yield about 0.90. TD Ratio performs the worst with an F1 of 0.75.

For UC2, it is crucial to minimize the number of false positives. A false positive is akin to crying wolf when there is no actual threat -- such erroneous alerts quickly lead to frustration and distrust among tool users. Consequently, we consider F0.5 to be the most important column to answer RQ1.

Considering F0.5 for UC2, we find that Code Health (B) performs the best (0.87) followed by the simple LoC-based baseline and SotA ML close behind at 0.83 and 0.82, respectively. We find that Code Health takes the first spot thanks to the highest Pr in the competition (0.89). As for UC1, we acknowledge the accuracy of the simple LoC baseline and corroborate the observation by Bertrand \textit{et al.}~\cite{bertrand_replication_2023}.

MS-MI and SonarQube perform substantially worse. Among these options, TD Time (E) comes out on top with an F0.5 of 0.68 but the corresponding Rc is only 0.50. MS-MI (D) yields a questionable F0.5 of 0.62. TD Ratio (G), on the other hand, is clearly a poor liability detector. Relying on SonarQube's Maintainability Indexes B--E to highlight unmaintainable files corresponds to an F0.5 of only 0.12 -- 7x and 6x worse than Code Health and MS-MI, respectively. We notice that SonarQube TD Time (E) performs slightly better than MS-MI, but it is not the primary metric for highlighting problematic files in SonarQube.

\begin{figure*}
    \centering
    \includegraphics[width=0.75\textwidth]{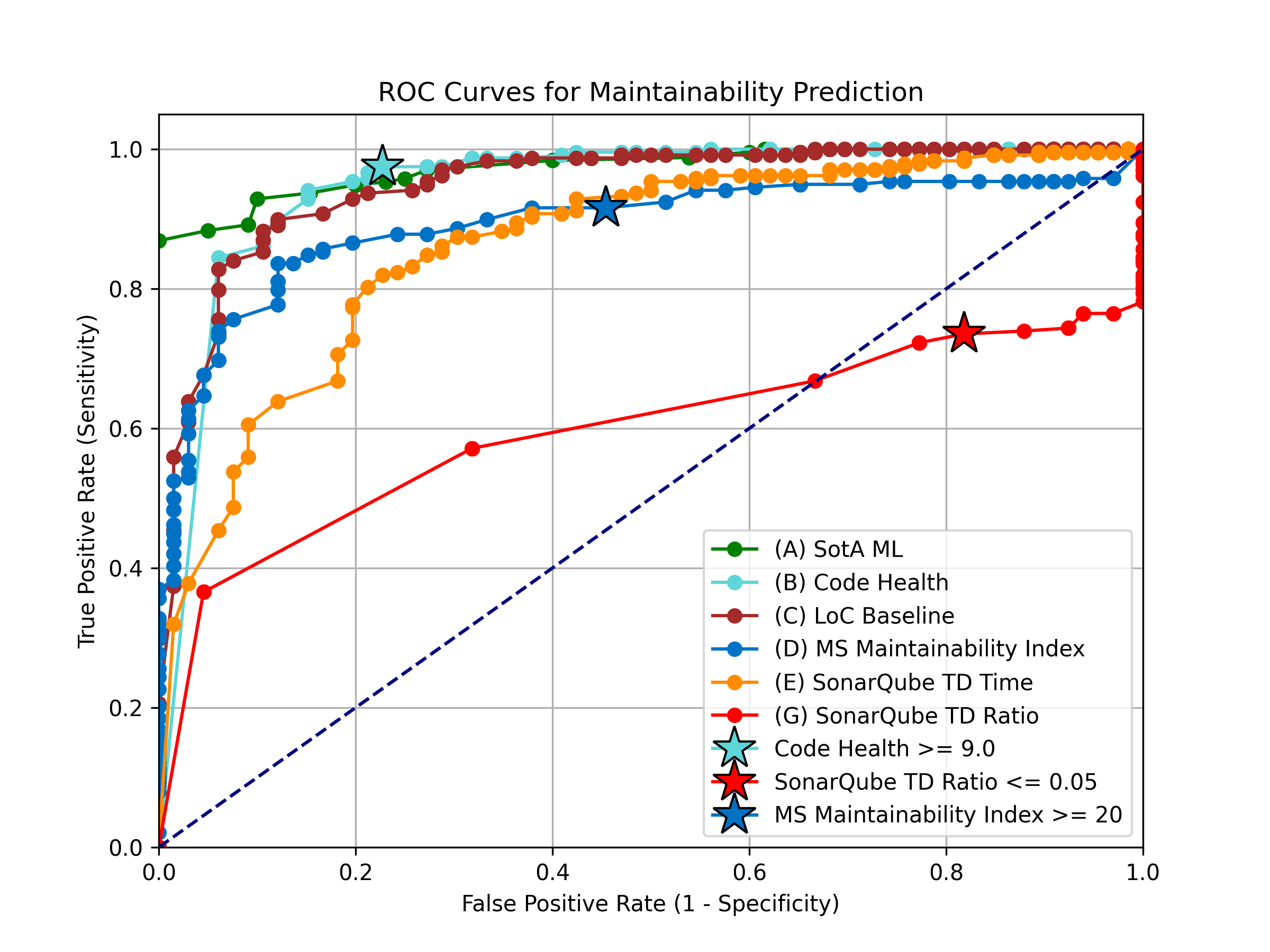}
    \caption{ROC curves for UC1 Maintainability Prediction.}
    \label{fig:roc_maint}
\end{figure*}

The stars in Figures~\ref{fig:roc_maint} and~\ref{fig:roc_liab} illustrate the tradeoffs between false positives and true positives that tool users experience. Focusing on UC2 Liability Prediction in Figure~\ref{fig:roc_liab}, CodeScene Code Health $<9.0$ yields a TPR (=recall) of 0.77 with a very low FPR of 0.025. MS-MI $<20$ results in a lower TPR (0.55) and a slightly higher FPR (0.084). For UC2 on MainData, SonarQube's default threshold of TD Ratio $>0.05$ performs worse than random chance, i.e., the corresponding star is under the dashed diagonal line. We investigate this phenomenon further in RQ2.

To conclude, LoC counting is a simple yet effective way to identify files that are hard to maintain. However, such an identification does not provide actionable input for a developer to improve the situation. Sure, the code should be split into smaller files, but beyond that, the developer is on her own. This is where the maintainability metrics based on code smells excel, i.e., they can provide actionable recommendations about what developers should do to improve the code. Code Health, SotA ML, and the LoC baseline all clearly outperform the average human expert -- but only the former includes clues for how to improve the code.


\begin{tcolorbox}[takeawaybox, title=RQ1: How do contemporary industrial tools’ maintainability assessments compare to state-of-the-art ML in matching human experts’ judgments?]
In terms of F1-scores for maintainability prediction, both CodeScene's Code Health (0.96) and Microsoft’s Maintainability Index (0.90) outperform the average human expert (0.88). CodeScene matches the performance of SotA ML (0.95). In contrast, SonarQube's Maintainability Rating (based on TD ratio) is notably less accurate (0.75) and should not be trusted for detecting maintainability issues.
\end{tcolorbox}

\subsection{RQ2: Underlying Predictive Powers}
\begin{figure*}
    \centering
    \includegraphics[width=0.75\textwidth]{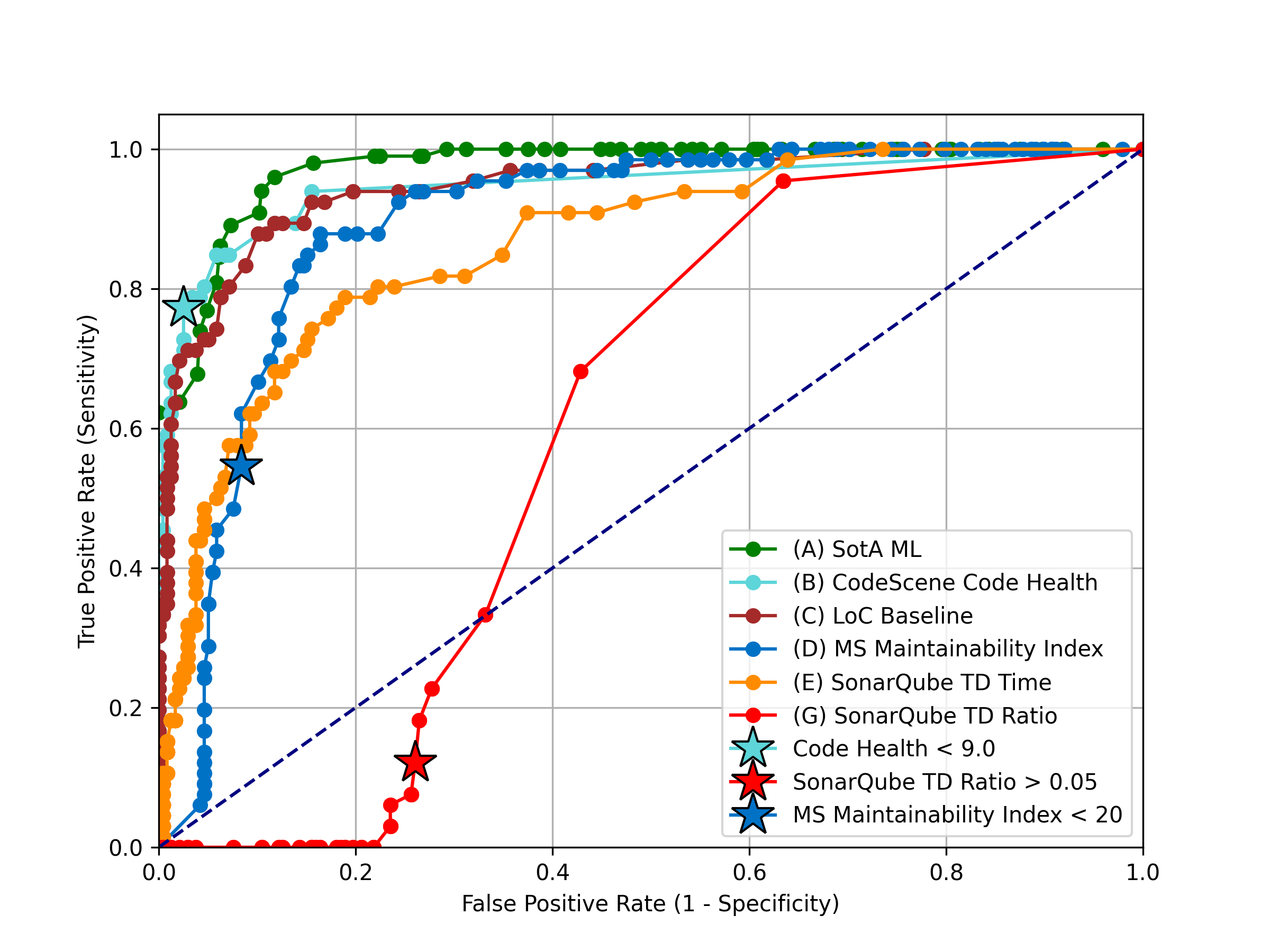}
    \caption{ROC curves for UC2 Liability Prediction.}
    \label{fig:roc_liab}
\end{figure*}

We find that, SotA ML (A), and Baseline LoC (C) visibly outperform the other prediction approaches. We recognize that the three approaches, and Code Health in particular, have threshold values that successfully limit the number of false positives -- the ROC curves in Figure~\ref{fig:roc_liab} contain many points to the very left.

The opposite is true for SonarQube TD Ratio (G), i.e., the primary maintainability metric the tool presents to its users. Its ROC curve shows that the SQALE method used to calculate the ratio between TD remediation time and the estimated total development time leads to many false positives. It is evident that the threshold for SonarQube's Maintainability Rating A does not reflect human experts' views. Our further investigations show that small files containing just a few Java statements often were flagged as highly problematic -- just a minor violation of a code convention in such a file can turn it into a major TD scapegoat.

\begin{figure*}
    \centering
    \includegraphics[width=0.99\textwidth]{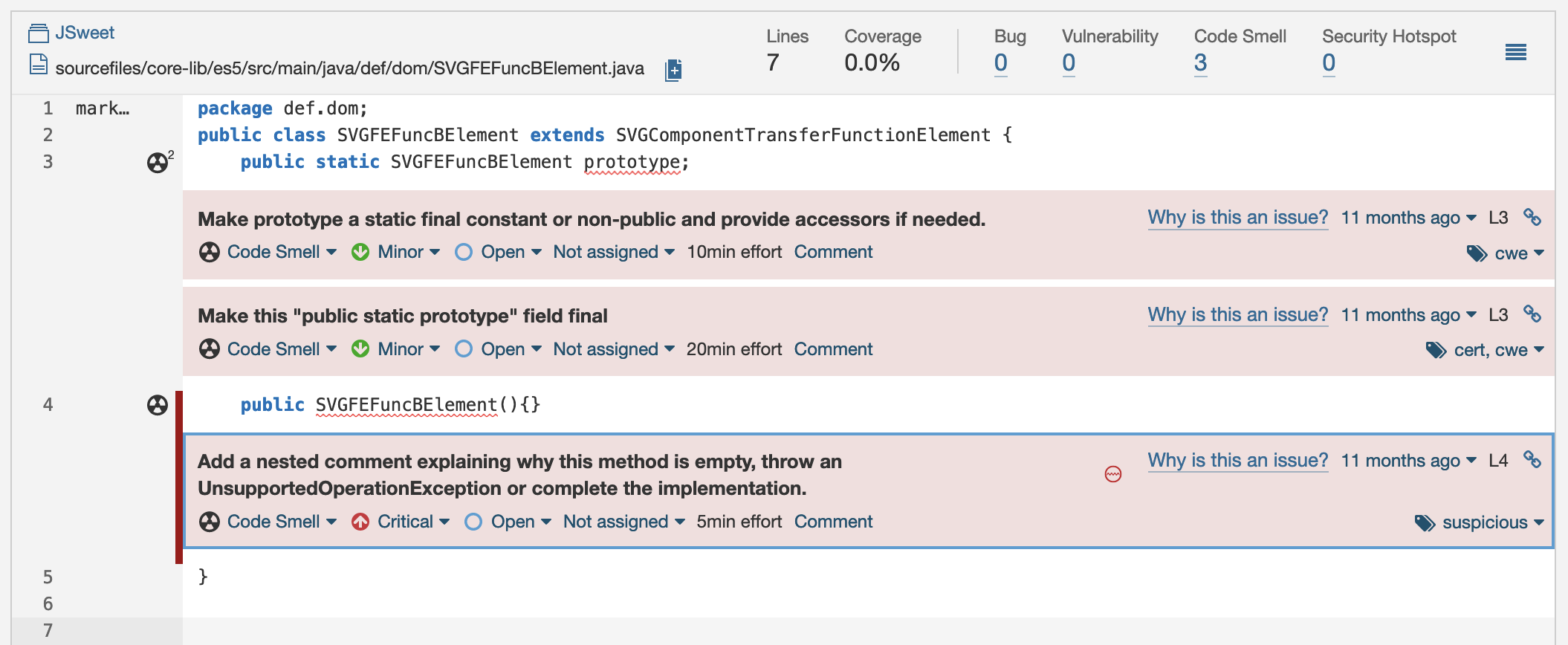}
    \caption{\texttt{SVGFEFuncBElement.java}, a very small file in JSweet. SonarQube yields a false positive, all other maintainability prediction approaches consider the file maintainable. The five lines of code contain three SonarQube code smells (expanded in pink) which translates into a TD Time of 35 min and a TD Ratio of 0.233, resulting in Maintainability Rating D.}
    \label{fig:sonar}
\end{figure*}

Figure~\ref{fig:sonar} illustrates an example where SonarQube's TD Ratio leads to false alarms. In the first example, the very small Java class \texttt{SVGFEFuncBElement} contains three SonarQube code smells:
\begin{enumerate}
    \item The field \texttt{prototype} should either be declared as a static final constant or declared as non-public with accessor methods.
    \item The field \texttt{prototype} should be declared final.
    \item The constructor \texttt{SVGFEFuncBELement} should not have an empty body.
\end{enumerate}

\begin{figure*}
    \centering
    \includegraphics[width=0.99\textwidth]{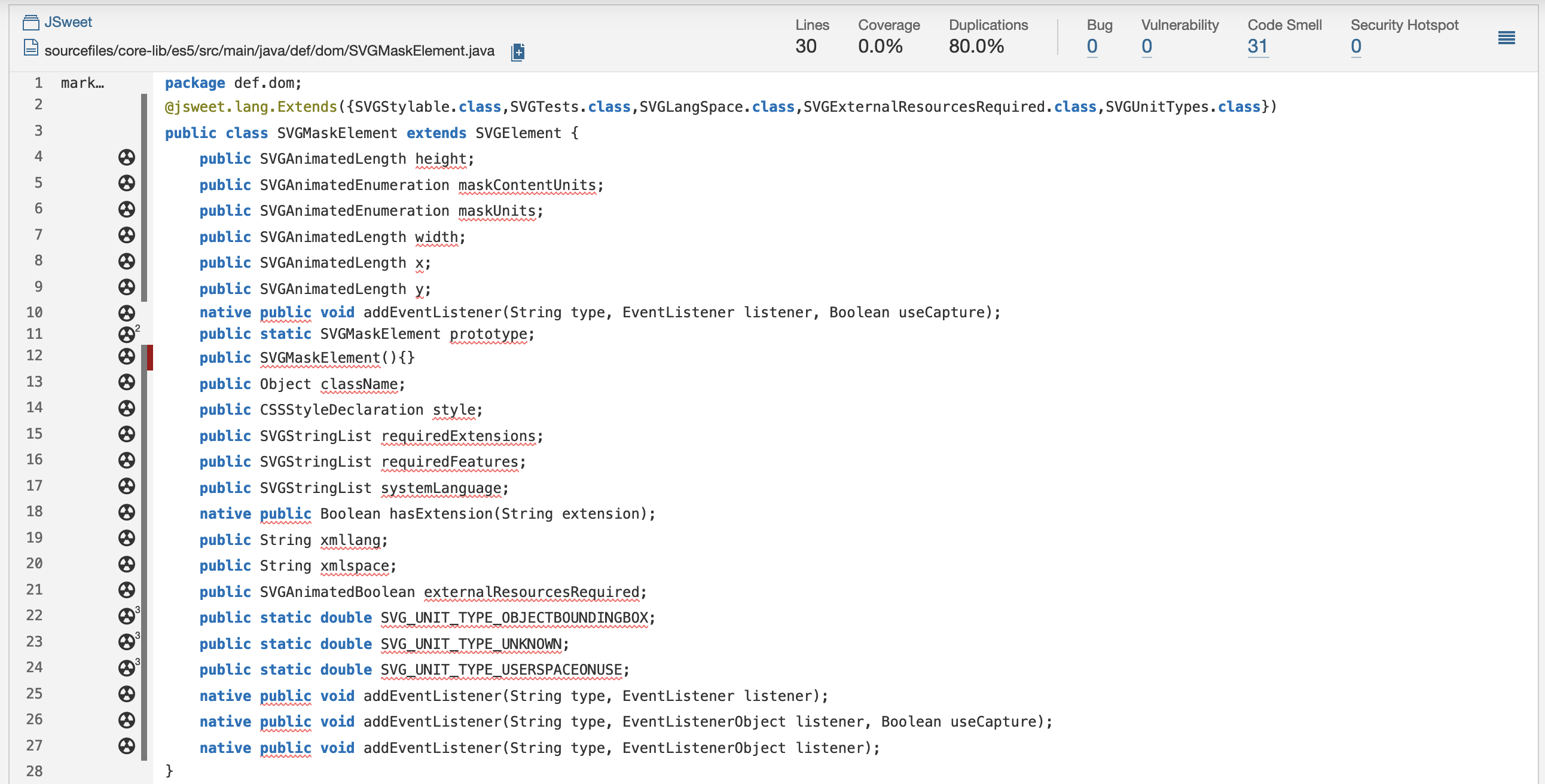}
    \caption{\texttt{SVGMaskElement.java}, a small file in JSweet with 31 SonarQube code smells. The file has a TD Time of 281 min and a TD Ratio of 0.335, resulting in Maintainability Rating D. According to SonarQube, this is the least maintainable file in the entire MainData. All non-SonarQube prediction approaches consider the file maintainable.}
    \label{fig:sonar2}
\end{figure*}

Figure~\ref{fig:sonar2} shows the class \texttt{SVGMaskElement}, another example of a false alarm. Almost every line contains a SonarQube code smell. In addition to the code smells described in the first example, SonarQube detects that several recommendations from the Java Language Specification are violated (the order of modifiers and that field names do not follow Java's naming conventions). Both examples have a SonarQube Maintainability Index D, but the human experts considered them maintainable -- as did all other maintainability prediction approaches investigated in this study.

Finally, we discuss the AUC scores of the different prediction approaches. Table~\ref{tab:res} presents the scores in the rightmost column. For (A)--(E) and (G) we calculate the areas under the ROC curves using the trapezoidal rule. For (F) in Table~\ref{tab:res}, we report the value from previous work~\cite{bertrand_replication_2023}. 

Our results show that Bertrand \textit{et al.}'s maintainability prediction results using AdaBoost (A) remain SotA for MainData with AUC 0.97. However, CodeScene Code Health (B) and a simple rule-based LoC baseline (C) come very close with AUC 0.95 -- outperforming both the other algorithmic prediction approaches and the average human expert (F with AUC 0.83). The accuracy of the LoC baselines evidently shows that human experts find large files hard to maintain. MS-MI (D) and TD Time (E) slightly outperform the average human expert. However, the TD Ratio (G) -- the metric leading to SonarQube's maintainability ratings A--E -- is by far the worst of the prediction approaches studied with AUC 0.60. 

\begin{tcolorbox}[takeawaybox, title=RQ2: What is the comparative predictive power of the tools' underlying maintainability metrics relative to state-of-the-art ML?]
Considering AUC, SotA ML (0.97) retains the leading position, closely followed by CodeScene's Code Health (0.95). Microsoft's Maintainability Index (0.89) and SonarQube's TD Time (0.86) perform worse but outperform the average human expert (0.83). At the bottom, SonarQube's TD Ratio (0.60) is very noisy with many false positives. 
\end{tcolorbox}

\section{Threats to Validity} \label{sec:threats}
\textbf{Construct validity.} The fundamental construct used in this study is maintainability. As with most qualities, it is a difficult subjective concept. As Schnappinger \textit{et al.} mention, aspects such as including commented-out code, hard-coded values, and poor comments are likely to influence the human assessment~\cite{schnappinger_defining_2020}. We are aware of the challenge but rely on the most rigorously annotated dataset available -- no other dataset offers such a reliable estimate of the construct. Note that seventy participants from 17 different companies were involved in the well-designed annotation process.

\textbf{External validity.} This is the major threat to the conclusions of this study. The 304 publicly available files of the MainData cannot possibly represent industry practice at large. Schnappinger \textit{et al.} argue that they sampled them from five open-source projects representing different application types, levels of complexity, and quality. Still, all five projects are implemented in Java. 

Nonetheless, as argued by Runeson \textit{et al.}, a useful benchmark can be seen as a set of selected cases that are meaningful to study for a question at hand, even if it is not representative of industrial practice as a whole~\cite{runeson_test_2008}. We strongly believe that MainData is a meaningful Java benchmark for seasoned developers' maintainability assessments. Moreover, after having worked with MainData during this study, we find that the files resemble legacy Java code as found in proprietary projects. While we believe our findings generalize to Java, other benchmark studies are needed for languages with other characteristics such as C, Rust, and Clojure.

\textbf{Internal validity.} We do not make causal claims beyond correlations between maintainability prediction approaches and a ground truth established by human experts. We have not identified any confounding variables, i.e., variables that affect both the prediction approaches and the human experts' annotations. Thus, we consider this category of threats minor.

\textbf{Conclusion validity.} We do not conduct any statistical inference testing in this study. Instead, we draw conclusions based on a set of established performance metrics. For RQ1, we primarily discuss the less common F0.5, which we believe is appropriate for UC2 in which false positives would annoy the user. For RQ2, we rely on AUC to draw conclusions, which is the recommended practice for binary classifiers~\cite{bradley_use_1997}. We could use smaller stepsizes when plotting RoC curves, which would yield more accurate AUC scores using the trapezoidal rule, but we consider our current estimates accurate enough.

\textbf{Reliability.} In software engineering research, discussions about reliability threats typically arise in the context of case study research~\cite{runeson_case_2012}. Such discussions focus on the extent to which the data and its analysis may be influenced by the individual researchers involved. We reuse the concept to discuss the influence of tool use. Different tools implement the MS-MI, but we rely only on MetricsTree. To mitigate this threat, we used MetricsTree's extracted CC, HV, and LoC to verify its internal implementation of MS-MI on a random sample of about 100 files.

\section{Implications for Research and Practice} \label{sec:impl}
We believe that our study has significant implications. Although several academic studies have raised concerns about the accuracy of SonarQube's output~\cite{saarimaki_accuracy_2019,baldassarre_diffuseness_2020,lenarduzzi_critical_2023}, it continues to be widely used in research to create ground truth quality labels. Despite many competing code analysis tools, SonarQube -- released in 2006 -- remains market-leading in the software industry. Our study further challenges this status quo and we highlight four primary implications in academic and practical domains.

First, SonarQube implements the SQALE to prioritize TD targets, including its method to use TD Time to calculate TD Ratio~\cite{letouzey_managing_2012}. Specifically, although SQALE is designed to prevent the misidentification of all large files as problematic, our analysis of MainData indicates that it predominantly flags small files as the main TD culprits. On the other hand, as shown by the accuracy of the LoC baseline, human experts indeed find large files to be troublesome. This finding also raises concerns about using SonarQube's TD Time per LoC as a maintainability metric, sometimes referred to as \textit{Technical Debt Density} (TDD), which has been increasingly popular in recent years~\cite{paudel_towards_2024,digkas_can_2022,zabardast_impact_2022}. We call for additional research to better accommodate file size in maintainability assessments. 

Second, our findings echo previous studies, showing that SonarQube's TD Time does not align well with human expert judgments. In contrast, CodeScene's Code Health demonstrates a higher predictive power on MainData (AUC 0.95 vs. AUC 0.86) and the corresponding number of false positives is substantially lower for UC2 (Pr 0.89 vs. Pr 0.75). Remarkably, even simpler metrics such as the MS-MI and the naïve LoC baseline outperform SonarQube's TD Time. Our analysis indicates that the aggregation of CodeScene's 25 high-level Java code smells aligns more closely with what human experts consider maintainability obstacles, unlike SonarQube's detection of hundreds of low-level code smells. A large portion of SonarQube rules belongs in the domain of linting tools, which apparently are less effective at detecting meaningful maintainability issues. Academic maintainability researchers should at least consider metrics beyond SonarQube output. To facilitate further research, we now offer Code Health scores for MainData~\cite{borg_ghost_2024}. A promising next step would be to analogously complement the Technical Debt Dataset with Code Health~\cite{lenarduzzi_technical_2019}.

Third, our study underscores the value of trustworthy benchmarks for maintainability research. The endeavor by Schnappinger \textit{et al.} to establish MainData with highly reliable human labels is commendable. Using MainData, we have revealed important insights into the predictive powers of contemporary industrial tools. However, as further discussed in Section~\ref{sec:threats}, MainData is restricted to Java -- and relatively outdated versions of this popular language. Academic research always calls for more comprehensive benchmarks, and we voice the same in this paper. We suggest exploring scalable methods for creating labeled maintainability datasets, potentially by integrating Schnappinger \textit{et al.}'s methodology with crowdsourcing approaches~\cite{sari_systematic_2019}.  

Fourth, simply counting LoC is a remarkably accurate identifier of unmaintainable files. On the other hand, as discussed in Section~\ref{sec:res_rq1}, LoC does not provide developers with a practical starting point for refactoring efforts. The same goes for the even more accurate predictions of SotA ML, i.e., a developer can not act on improving a file based on the low-level code metrics used by an ensemble learner. Instead, we argue that more research should be focused on the actionability of tool outputs. In line with the recent roadmap for TD research~\cite{avgeriou_technical_2023}, we are strong advocates of increasing the level of abstraction in maintainability management to better align with human judgments. In this light, we posit that CodeScene's high-level code smells are more actionable than the low-level counterparts detected by SonarQube.

\section{Conclusion} \label{sec:conc}
Working with trustworthy metrics is a cornerstone in software maintenance. In this study, we used a highly reliable maintainability dataset with human labels to benchmark three metrics employed by contemporary industrial tools. Furthermore, we compared the results to \textit{State-of-the-Art Machine Learning} (SotA ML) models for predicting binary maintainability at the file level.

First, we compared the prediction performance that end users experience under the default configurations of these tools. Our findings showed that CodeScene matched SotA ML in maintainability judgments and outperformed the average human expert. Microsoft’s Maintainability Index demonstrated a moderate level of accuracy. In contrast, SonarQube was notably less accurate and should not be trusted for detecting maintainability issues. 

Second, we investigated the predictive power of the tools' underlying maintainability metrics compared to SotA ML. SotA ML maintained the leading position with CodeScene’s Code Health marginally less accurate. Microsoft’s Maintainability Index and SonarQube’s Technical Debt Time performed worse but still exceeded the accuracy of the average human expert. At the bottom, SonarQube’s Technical Debt Ratio is very noisy and generates many false positives, i.e., ghost echoes.

We conclude that many of the low-level code smells used in SonarQube do not reflect what humans perceive as maintainability issues. Low-level details such as breaking naming rules and violating coding conventions are better suited for linting tools rather than maintainability tools. Regrettably, several previous studies relied on SonarQube output for establishing ground truth labels for maintainability and technical debt research. For future studies, we recommend using more reliable metrics, and suggest reevaluating previous findings with Code Health to mitigate related threats to validity.

\section*{Acknowledgment}
This work was partly funded by the NextG2Com Competence Centre -- Next-Generation Communication and Computing Infrastructures and Applications -- under the Vinnova grant number 2023-00541.

\section*{Conflicts of Interest}
Tornhill is the founder of CodeScene and the main developer of the Code Health metric. Borg leads the company's research activities. Ezzouhri interned at CodeScene in 2024.

\bibliographystyle{IEEEtran}
\bibliography{IEEEabrv, codered}

\end{document}